\documentclass[preprint]{aastex}

\begin{document}

\title{Measurement of the magnetic field direction in the \\
NGC2024-FIR5 protostellar outflow}
\author{J.S. Greaves and W.S. Holland}
\affil{Joint Astronomy Centre, 660 N. A`oh\={o}k\={u} Place, 
University Park, Hilo, HI~96720 
\\
{\em j.greaves@jach.hawaii.edu, w.holland@jach.hawaii.edu}}
\and
\author{D. Ward-Thompson}
\affil{Department of Physics and Astronomy, Cardiff University, 5 The
Parade, Cardiff CF2 3YB, UK 
\\
{\em D.Ward-Thompson@astro.cf.ac.uk}}

\begin{abstract} 

Molecular outflows from young protostars are widely believed to be collimated by
magnetic fields, but there has been little observational evidence to support this
hypothesis. Using the new technique of millimetre-wavelength spectropolarimetry,
we demonstrate the existence of a magnetic field in the NGC2024-FIR5 outflow
lobe. The 1.3 mm J=2--1 transition of carbon monoxide (CO) is polarized at a
level of approximately 1 \%, in a direction within 10--15$^{\circ}$ of the
outflow axis.  This agrees with theoretical models where the magnetic field
channels the outflowing gas, and shows that the process can be effective as
far as 0.1 pc from the protostar. 

\end{abstract}

\keywords{ISM: magnetic fields -- ISM: jets and outflows -- polarization --
stars: formation}

\lefthead{Greaves et al.}
\righthead{Magnetized outflow in NGC2024}

\section{Introduction}

Many models predict that the outflows from young protostars are
magnetically collimated (e.g. Uchida \& Shibata 1985; Pudritz 1986; Shu
et al. 1994;  Fiege \& Henricksen 1996). The collimation may take place
within a few stellar radii, or be maintained at much larger distances,
for example by alignment of the magnetic field with streamlines in the
extended protostellar envelope. These theories have previously been very
difficult to test, as the outflow emission is weak compared to that of
the ambient cloud. Detection of the Zeeman effect in molecular lines is
thus very difficult, and the more sensitive technique of mapping of
polarized thermal dust emission (e.g. Hildebrand et al. 1996) cannot be
used, as there is no velocity information to separate the envelope and
outflow.

This problem can now be addressed with spectropolarimetry in the
millimetre-wave regime, a recently developed technique (e.g. Greaves et al.
1999) utilising the Goldreich-Kylafis effect. In the presence of a magnetic
field, Zeeman splitting of the rotational levels of molecules occurs, and
rotational transitions can then be polarized, due to imbalances in the
magnetic sublevel populations (Goldreich \& Kylafis 1981). The percentage
polarization is sensitive to optical depth and excitation conditions, since
both polarizations saturate if $\tau \gg 1$, and the sub-level populations
equalise if collisional transition rates significantly exceed radiative
rates. However, with suitably chosen molecular lines, spectropolarimetry can
be used to map magnetic fields {\it as a function of velocity}. There is an
ambiguity in the field direction, since the polarization direction may be
either parallel or perpendicular to the net field in the plane of the sky,
but this can be resolved for simple velocity structures (Kylafis 1983).

Detections of magnetic fields using this technique have previously been
published by Glenn et al. (1997) for the envelope of the evolved star
IRC+10216, and by Greaves et al. (1999) for two molecular clouds and the
`2 pc ring' around the Galactic Centre. These are all bright sources and
the deepest observations so far are those of Girart, Crutcher \& Rao
(1999), who used the BIMA array to map the environment of the
NGC1333-IRAS4A protostellar system. Surprisingly, they did not find a
field clearly aligned with the outflow axis, but did detect strong
polarization (2--10\%) in a region within the flow where the magnetic
field appears to be deflecting the gas.

In this {\it Letter}, we present results of a similar experiment for
the molecular outflow from the vicinity of the massive young source
NGC2024-FIR5. This outflow has been extensively mapped (e.g. Richer
et al. 1992) and extends several arcmin to the south, but has
apparently no northern lobe. We have used the James Clerk Maxwell
Telescope to search for polarized CO J=2--1 rotational emission (at
1.3 mm wavelength) in the bright, highly collimated, red-shifted
lobe. Polarization was detected at a low level ($\approx$ 1~\%), and
the results are discussed in the context of the outflow structure.

\section{Observations}

The observations were made on 10 December 1996, using the single-polarization
heterodyne receiver A2 (Davies et al. 1992) at the JCMT, located on Mauna
Kea, Hawaii. The backend was a digital auto-correlation spectrometer, set to
a resolution of 0.156 MHz. Standard observing techniques were used, with
position-switching 0.5$^{\circ}$ west for sky subtraction, and three-load
calibration (ambient, sky and cold load). The polarimeter module, consisting
of a rotating half-wave plate (Murray et al. 1992), was mounted externally in
front of the receiver, and the wave plate was `stepped' at intervals of
22.5$^{\circ}$ around a complete circle (one waveplate cycle), to produce a
set of 16 spectra. The half-wave plate rotates the source plane of
polarization with respect to that accepted by the single-polarization
receiver, and modulations in the CO spectra were subsequently analysed as a
function of waveplate angle. The reduction techniques were developed after
the observations were completed, and are described below and in more detail
by Greaves et al. (1999). 

Two positions were observed, centred on the positions shown in Figure~1.  
The integrated intensities in the outflow line wings, as a function of
waveplate position, were fitted for the polarization parameters using a
least-squares procedure (Nartallo 1995), which gives the Stokes parameters
Q and U of the polarization, in the receiver frame of reference.
Half-cycles were fitted, each containing 8 spectra, and points were
allowed to be dropped to improve the fits, although only 8~\% of the final
data was removed in this way. The noise appears to be dominated by a few
anomalous signals (probably produced by calibration errors or brief sky
changes), and removing these data decreases the reduced-$\chi^2$ of the
fits by a factor of two. If these poorer data are left in, the final
position angles change by about 10--15$^{\circ}$.

An instrumental polarization (IP) was then subtracted, measured
using the same technique on the 1.3 mm continuum emission of Saturn,
which was assumed to be unpolarized. The IP was found to be $0.5 \pm
0.15$~\% at a position angle of $169 \pm 8^{\circ}$ in az-el
co-ordinates, and is thought to be dominated by the vertical weave
of the JCMT `windblind'. The IP measurement used was calculated over
the wing velocities, but is the same within the errors for the whole
passband used (approximately 300 km s$^{-1}$). The IP-subtracted
values were then corrected for parallactic angle and a fixed
rotation due to the instrument position in the receiver cabin, and
coadded to give the final results listed in Table~1. Additionally,
the data were reduced by the same algorithms but subtracting spectra
taken 45$^{\circ}$ apart, to produce Stokes spectra Q(v) and U(v).
The results for the brighter outflow position are shown in Figure~2,
after converting to p(v),$\theta$(v), the polarization percentage
and direction (measured anti-clockwise from +Declination).

\section{Results}

The two positions observed are a bright peak in the red-shifted outflow,
and a more generic point 40 arcsec to the south. After 5 and 4 waveplate
cycles respectively (2560 and 2048 seconds of integration), the rms noise
levels are $\approx 20$ mK per rebinned 2 km s$^{-1}$ spectral channel,
compared to the line wing brightness of $\sim$ 5 K. The integrated
intensities in the wings were approximately 60 and 30 K km s$^{-1}$, in
the T$_A^*$ antenna temperature scale (corrected for atmospheric opacity
but not the main-beam efficiency, which was approximately 0.7). The
fainter position is about three times lower in integrated intensity than
any of our previously published data (Greaves et al. 1999).

Polarization was detected at both positions, with similar levels of around
1~\% (Table~1). The significance level of the data is high, with detections
of 3.5 and 4.8 sigma. Using the statistical methods of Clarke \& Stewart
(1986), these detections are real at the 99.9~\% confidence level. No special
procedures were used to enhance the signal-to-noise, although one fit was
eliminated for each position, out of the ten and eight originally obtained.
One of these discrepant fits was clearly associated with a change in
intensity (most likely a sky fluctuation), while the other was not explained
but the Stokes parameters were out of phase by $\sim 80^{\circ}$ compared to
the final result.

To confirm that the results are real, a further short observation of the
northern position was made on 13 August 2000. Polarization spectra from a
single waveplate cycle indicated p,$\theta$ of $1.8 \pm 0.5$~\%,$-5 \pm
8^{\circ}$. Although conditions allowed only this short test, the result
is in good agreement with the 1996 data. Significantly, the observing
parameters were completely different, including a new receiver, wider
spectrometer setting, different planet (Uranus) used for the instrumental
polarization measurement, and a parallactic angle higher by 100$^{\circ}$.
A further `null' test for instrument systematics is to observe a source
where little polarization is expected, for example a very optically thick
line. This test was performed in July 1996, during polarimetry of the CO
J=3--2 emission towards the Galactic Centre (cf. Greaves et al. 1999), and
p $\sim 0.1 \pm 0.3$~\% was measured in a very bright position in the
south of the `2 pc ring'. This confirms that polarization is not always
detected and strengthens the case for low instrumental residuals.

\subsection{Outflow polarization}

The polarization position angles are --6$^{\circ}$ and +21$^{\circ}$, and are
well aligned with the outflow axis. We estimate the flow orientation to be
5$^{\circ}$ east of north on large scales, although the angle may be higher
nearer to the origin, presumed to be FIR5 (marked in the figure by a square
symbol) or a fainter source nearby (Chernin 1996). Thus the polarization is
within about 10$^{\circ}$ and 15$^{\circ}$ of the flow axis at the north and
south points, respectively. There may be a slight bending of the polarization
coinciding with local fluctuations in flow direction (Figure~1), but this is
at the limits of the errors.

The percentage polarization is in very good agreement with theoretical
values.  Deguchi \& Watson (1984) have modelled CO polarization in sources
with velocity gradients, and for the outflow optical depth and density
conditions ($\tau$(CO) $\sim$ 5, n(H$_2$) $\sim$ 10$^3$ cm$^{-3}$, Richer et
al. 1989, 1992), they predict p values up to about 1.0 \%, depending on
viewing angle (their Figures 2 and 5).  The observed values are at this
level, consistent with the calculations if the outflow is close to the plane
of the sky, as seems likely from the large extent and high apparant
collimation.

The polarized spectra for the north position are shown in Figure~2. The
errors for single spectral channels are considerable (averaging $\pm 0.6$~\%
over 2~km~s$^{-1}$), and thus only broad conclusions can be drawn, but the
percentage and direction in the red wing appear to be approximately constant.
The mean values are 1.0~\% and --9$^{\circ}$, in good agreement with the
results from the fits (Table~1), confirming that the two analysis methods
agree. The polarization spectra for the South position are considerably
noisier per channel, as the wing intensity is a factor of two less, and these
spectra are not shown.

\section{Discussion and Conclusions}

The observed polarization direction is a direct diagnostic of the
orientation of the magnetic field in the plane of the sky. Kylafis (1983)
has shown that the polarization will be either perpendicular or parallel
to the field direction, even for quite low field strengths (B $\sim$ 1
$\mu$G). Further, if the outflow is assumed to have a one-dimensional
velocity field, it is possible to distinguish between these two
orientations. For large $\tau$(CO), the polarization will generally be
seen parallel to the field, unless the field is at a large angle to the
velocity gradient (Kylafis 1983). The criterion for the field and
polarization to be parallel is cos$^2 \alpha > 1/3$ ($\alpha <
54.7^{\circ}$), where $\alpha$ is the angle between the velocity gradient
and the field.

For NGC2024, Richer et al. (1992) have found that the flow is accelerated,
with the fastest moving material further from the star, so we assume that the
dominant velocity gradient is along the outflow axis. Then the most probable
interpretation of the data is that the magnetic field lies close to the flow
axis, i.e. $\alpha \sim 0^{\circ}$, and the polarization will be produced
near the flow axis, as observed.  It is unlikely that the p vectors are
perpendicular to the field, as then the field would lie across the flow axis
($\alpha \sim 90^{\circ}$), and this would impede the flow by channeling
ionized molecules sideways.

Magnetic channeling of the outflow is not implausible on energy grounds.
For a gas cylinder, the magnetic energy is given by $B^2 R^2 / 8$ for a
unit length (e.g. Chandrasekhar \& Fermi 1953), and the outflow kinetic
energy will be $M v^2 / 2$. Adopting an outflow radius (cf. Figure~1) of
$R \sim 0.03$~pc or $1 \times 10^{17}$~cm, the mass per unit length is
$M \sim 3 \times 10^{14}$ g, and a lower limit on the bulk outflow speed
is $v \sim 10$~km s$^{-1}$ (cf. Figure~2). This gives a required
magnetic field strength, for the two energies to be equivalent, of $\geq
350 \mu{\rm G}$. However, since the magnetic field affects only {\it
orthogonal} flow, then the energy required is reduced when the field and
flow are nearly aligned. Adopting an outflow opening angle of $\pm
5-10^{\circ}$ (cf. Fig.~ 4 of Richer et al. 1992), the cross-wise
velocity is only 1--2~km~s$^{-1}$ and B $\geq 35-70 \mu{\rm G}$ should
provide sufficient energy to keep the flow well collimated. This is not
excessive, given that Crutcher et al. (1999) have measured a lower limit
to the line-of-sight field component of 80 $\mu{\rm G}$, in the main
cloud emission near our North point.

In summary, the spectropolarimetry observations have shown that the outflow
gas is in fact magnetized, and that the net field direction is aligned within
a few degrees of the flow axis, as predicted in most models. If the magnetic
field is channeling the outflowing gas, this process must be effective at
quite large distances from the star (1 arcmin, or about 0.1 pc at a distance
of 415 pc). We see little sign of `magnetic deflections' as in the
NGC1333-IRAS4A outflow observed in CO J=2--1 spectropolarimetry by Girart et
al (1999), using the BIMA array with a $9 \times 6''$ synthesized beam. The
NGC2024 outflow is relatively straight, although the small differences in
our North and South polarization directions do appear to follow slight bends
in the flow at the limits of the errors.

\acknowledgements

We would like to thank Per Friberg, Bill Dent, Sye Murray and Ramon
Nartallo for vital assistance with the polarimeter hardware and software,
Ramprasad Rao for some very useful discussions, and an anonymous referee
for helpful comments. The JCMT is operated by the Joint Astronomy Centre,
on behalf of the UK Particle Physics and Astronomy Research Council, the
Netherlands Organisation for Pure Research, and the National Research
Council of Canada.

\newpage

\begin{center}
{\bf Figure Captions}
\end{center}

\figcaption[f1.ps] {Map of the NGC2024-FIR5 molecular outflow, with
polarization data superimposed. The map data are for the J=3--2 (0.8 mm)
transition of CO observed with a 14$''$ FWHM beam size and taken from the
JCMT archive. Contour intervals are approximately 0.25 K (T$_A^*$ antenna
temperature scale) for the mean wing intensity in the range v$_{lsr}$ =
+18 to +40 km s$^{-1}$. The white square marks the FIR5 source and
positions are given in 1950 co-ordinates. The polarization vectors are
plotted from the Table~1 results, within beam-sized circles of 21$''$ at
1.3 mm. }

\figcaption[f2.ps] {Polarization spectra for the North position in the
outflow. The top and bottom panels show, respectively, the percentage and
direction of polarization, with the intensity spectrum superimposed. The
polarization spectra are binned by 10 channels compared to the intensity
spectra and are corrected for bias (cf. Table 1). Errors shown are the
standard deviation per velocity channel.}

\newpage


\begin{deluxetable}{lcccc}
\tablenum{1}
\tablewidth{0pt}
\tablecaption{Line polarization results.}
\tablehead{
\colhead{Position} & \colhead{RA,Dec (1950)} & \colhead{v$_{lsr}$ 
(km s$^{-1}$}) & \colhead{p (\%)} & \colhead{$\theta$ ($^{\circ}$)} }
\startdata
North & $05h\; 39m\; 12.1s, -01^{\circ}\; 57'\; 24''$ & 18--30 & 0.91 $\pm$ 0.26 & --6 $\pm$ 8\nl 
South & $05h\; 39m\; 12.1s, -01^{\circ}\; 58'\; 04''$ & 20--34 & 1.30 $\pm$ 0.27 &  21 $\pm$ 6\nl 
\enddata
\tablecomments{Polarization results are for the range of outflow
velocities given, and percentages are corrected for the positive-definite 
bias due to squaring the Stokes parameters (Wardle \& Kronberg 1974).
Positions are at $\delta$RA,$\delta$Dec = (--8,--24) and (--8,--64) arcsec
relative to the map origin in Figure~1. }
\end{deluxetable}


\newpage

\begin{figure}
\setlength{\unitlength}{1mm}
\begin{picture}(80,220)
\includegraphics{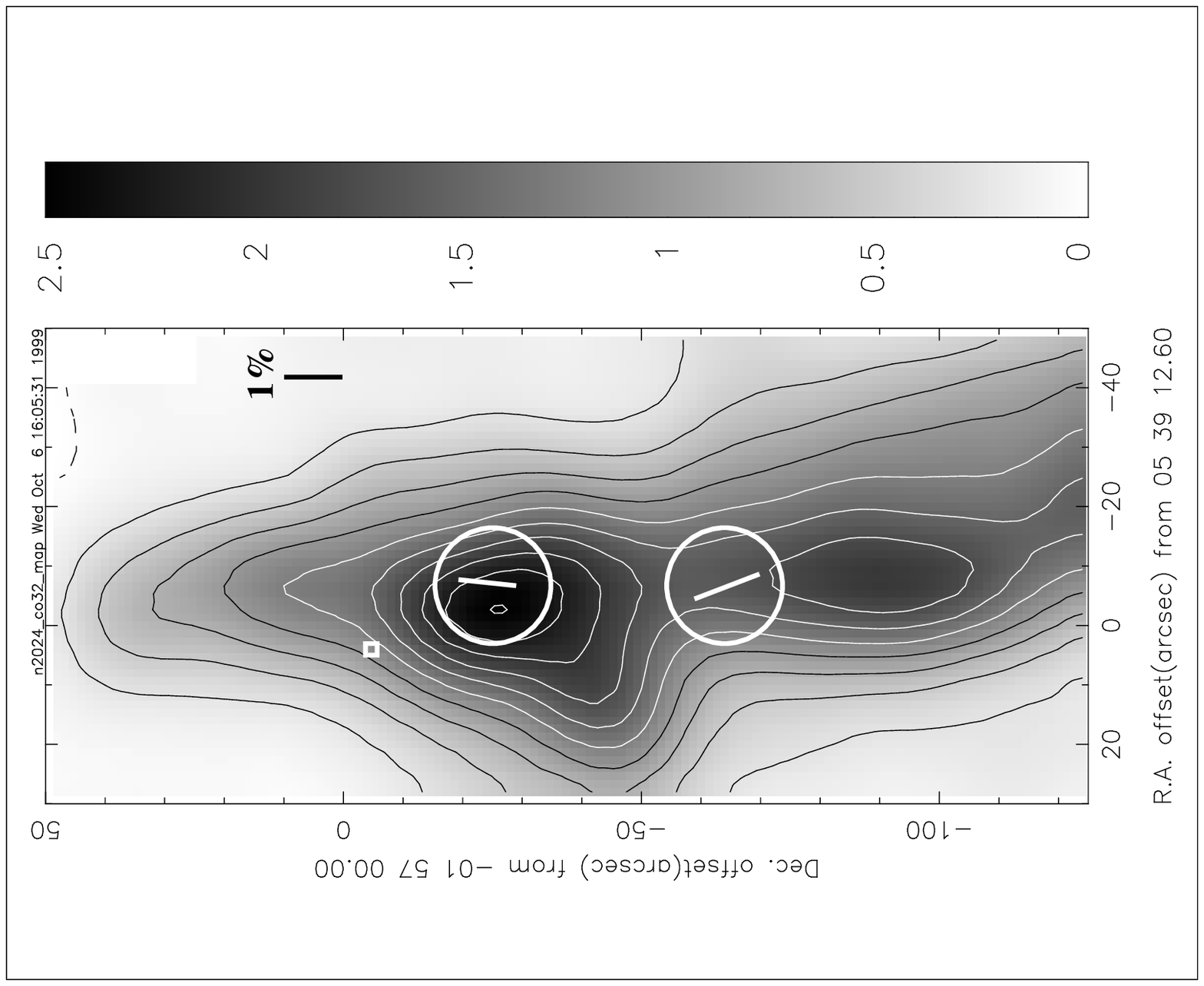}
\end{picture}
\end{figure}

\begin{figure}
\setlength{\unitlength}{1mm}
\begin{picture}(80,120)
\includegraphics{f2.eps}
\end{picture}
\end{figure}

\end{document}